\documentclass[aps,twocolumn,amssymb]{revtex4}
\usepackage{amssymb}
\usepackage{graphicx}
\usepackage{color}
\usepackage{textcomp}
\usepackage{ulem}
\begin{document}

\title{Correlations and fluctuations in a magnetized PNJL model with and without inverse magnetic catalysis effect}
\author{Shijun Mao}
 \email{maoshijun@mail.xjtu.edu.cn}
\affiliation{School of Science, Xi'an Jiaotong University, Xi'an, Shaanxi 710049, China}

\begin{abstract}
The correlation $\chi^{BQ}_{11}$ and quadratic fluctuations $\chi^B_2,\ \chi^Q_2,\ \chi^T_2$ of baryon number $B$, electric charge $Q$ and temperature $T$ are investigated in a two-flavor Polyakov loop extended Nambu-Jona-Lasinio (PNJL) model at finite temperature and magnetic field. The inverse magnetic catalysis (IMC) effect is introduced through the magnetic field dependent parameters $G(eB)$ or $T_0(eB)$, and we make comparison of the results in the cases with and without IMC effect. With nonvanishing magnetic field, the correlation $\chi^{BQ}_{11}$ and fluctuations $\chi^B_2,\ \chi^Q_2,\ \chi^T_2$ increase with temperature, and then show the peak around the pseudocritical temperatures of chiral restoration and deconfinement phase transitions in the cases with and without the IMC effect. The correlation and fluctuations along the phase transition line under external magnetic field are characterized by the scaled correlation ${\hat {\chi}}_{11}^{BQ}=\frac{\chi_{11}^{BQ}(eB,T_{pc}^c(eB))}{\chi_{11}^{BQ}(eB=0,T_{pc}^c(eB=0))}$ and scaled fluctuations ${\hat {\chi}}_2^{B(Q,T)}=\frac{\chi_2^{B(Q,T)}(eB,T_{pc}^c(eB))}{\chi_2^{B(Q,T)}(eB=0,T_{pc}^c(eB=0))}$ at the pseudocritical temperature $T_{pc}^c$ of chiral restoration phase transition. ${\hat {\chi}}_{11}^{BQ},\ {\hat {\chi}}_2^{B}$, and ${\hat {\chi}}_2^{Q}$ increase with magnetic fields, and the inclusion of IMC effect leads to some enhancement in their values. However, ${\hat {\chi}}_2^{T}$ is altered by the IMC effect. Without IMC effect, ${\hat \chi}^T_2$ slightly increases and then decreases with magnetic fields. Taking into account of the IMC effect by $G(eB)$, ${\hat \chi}^T_2$ monotonically increases with magnetic fields, and by $T_0(eB)$, it is a nonmonotonic function of magnetic field.
\end{abstract}

\date{\today}

\maketitle

\section{Introduction}
Motivated by the strong magnetic field in the core of compact stars and in the initial stage of relativistic heavy ion collisions, the study on Quantum Chromodynamics (QCD) phase structure under external electromagnetic fields has attracted much attention~\cite{review0,review1,review2,review3,review4,review5,lattice1,lattice2,lattice4,lattice5,lattice6,lattice7,lattice9,lattice8,fukushima,mao,kamikado,bf1,bf13,bf2,bf3,bf5,bf51,bf52,bf8,bf9,bf11,db1,db2,db3,db5,db6,pnjl1,pnjl2,pnjl3,pnjl4,pqm,ferr1,ferr2,mhuang,meimao1,t0effect,meihuangmao,t0effectmao}. 

The LQCD simulations~\cite{lattice1,lattice2,lattice4,lattice5,lattice6,lattice7,lattice9} observe the inverse magnetic catalysis (IMC) phenomena of u and d quarks, the decreasing chiral condensates near the pseudocritical temperature $T^c_{pc}$ of chiral restoration phase transition with increasing magnetic field, and the reduction of the pseudocritical temperature $T^c_{pc}$ under external magnetic field. Meanwhile, it is reported that the renormalized Polyakov loop increases with magnetic fields and the transition temperature of deconfinement decreases as the magnetic field grows~\cite{lattice1,lattice2,lattice4,lattice5,lattice9}. On analytical side, in the presence of a uniform external magnetic field ${\bold B} =B{\it {{\bold e}_z}}$, the energy dispersion of quarks takes the form $E=\sqrt{p_z^2+2|Q_qB|l+m^2}$ with the momentum $p_z$ along the direction of magnetic field and the Landau level $l=0,1,2,...$~\cite{landau}. Due to this fermion dimension reduction, almost all model calculations at mean field level present the magnetic catalysis (MC) effect, the enhancement of chiral condensates in the whole temperature region and the increasing pseudocritical temperatures for chiral restoration and deconfinement phase transitions under external magnetic field, see reviews~\cite{review0,review1,review2,review4,review5} and the references therein. How to explain the inverse magnetic catalysis phenomena of chiral condensates and the reduction of pseudocritical temperatures for chiral restoration and deconfinement phase transitions is open questions. Many scenarios are proposed~\cite{fukushima,mao,kamikado,bf1,lattice9,bf13,bf2,bf3,bf5,bf51,bf52,bf8,bf9,bf11,db1,db2,db3,db5,db6,pnjl1,pnjl2,pnjl3,pnjl4,pqm,ferr1,ferr2,mhuang,meimao1,t0effect,meihuangmao,t0effectmao}, such as magnetic inhibition of mesons, sphalerons, gluon screening effect, weakening of strong coupling, and anomalous magnetic moment. In our current paper, the IMC (MC) generally refers to the decreasing (increasing) behavior of physical quantities under external magnetic field.

In addition to the QCD phase diagram, the thermodynamical properties of QCD matter are influenced by the external magnetic field. Among the thermodynamic quantities, the correlations and fluctuations of the conserved charges are accessible in both theoretical calculations and experimental measurements, and serve as the useful probes to study the QCD phase transitions, such as to identify the critical end point (CEP) of QCD phase diagram in the temperature-baryon chemical potential plane~\cite{xwork24,xwork25,xwork26,xwork27,xwork28,xwork29,xwork30}. However, they are much less explored at the chiral restoration phase transition (crossover) with finite temperature and vanishing chemical potential under external magnetic field. The analytical investigations at vanishing chemical potential, finite temperature and magnetic field~\cite{dingref31,dingref33,dingref34,dingref35,dingref36} have been conducted in frame of the hadron resonance gas model, Polyakov loop extended Nambu-Jona-Lasinio (PNJL) model and Polyakov loop extended chiral quark model. People suggest to use the correlation and/or fluctuation as the magnetometer of QCD~\cite{dingref31,ding2022,dingprl2024}. It should be mentioned that the peak structure of quadratic fluctuations around the pseudocritical temperature $T^c_{pc}$ in LQCD calculations~\cite{ding2022} can not be realized in these analytical calculations. Meanwhile, the IMC effect is not well considered in the previous analytical studies~\cite{dingref31,dingref33,dingref34,dingref35,dingref36}.

In our current paper, the correlation $\chi^{BQ}_{11}$ and quadratic fluctuations $\chi^B_2,\ \chi^Q_2,\ \chi^T_2$ of baryon number $B$, electric charge $Q$ and temperature $T$ are investigated in a two-flavor PNJL model at finite temperature and magnetic field. The IMC effect is introduced through two methods, the magnetic field dependent coupling between quarks $G(eB)$ and magnetic field dependent interaction between quarks and Polyakov loop $T_0(eB)$, respectively. The comparison among the results in the cases with and without IMC effect are made.

The paper is organized as follows. Section \ref{2fframe} introduces the two-flavor magnetized PNJL model and the definition of correlation $\chi^{BQ}_{11}$ and fluctuations $\chi^B_2,\ \chi^Q_2,\ \chi^T_2$. Section \ref{results} discusses the numerical results of correlation and quadratic fluctuations at finite temperature and magnetic field in the cases without IMC effect and with IMC effect. Finally, we give the summary in Sec.\ref{summary}.

\section{theoretical framework}
\label{2fframe}

The two-flavor PNJL model under external magnetic field is defined through the Lagrangian density~\cite{pnjl5,pnjl6,pnjl7,pnjl8,pnjl9,pnjl10,pnjl12,t044,t041},
\begin{eqnarray}
\label{pnjl}
{\cal L} &=& \bar\psi(i\gamma_\mu D^\mu-\hat{m}_0)\psi\\
 &+& {G\over 2}\left[\left(\bar\psi\psi\right)^2 + \left(\bar\psi i \gamma_5 {\bf \tau}\psi\right)^2 \right]-{\cal U}(\Phi,\bar\Phi).\nonumber
\end{eqnarray}
The covariant derivative $D^\mu=\partial^\mu+i Q_f A^\mu-i {\cal A}^\mu$ couples quarks to the two external fields, the magnetic field ${\bf B}=\nabla\times{\bf A}$, and the temporal gluon field  ${\cal A}^\mu=\delta^\mu_0 {\cal A}^0$ with ${\cal A}^0=g{\cal A}^0_a \lambda_a/2=-i{\cal A}_4$ in Euclidean space. The gauge coupling $g$ is combined with the SU(3) gauge field ${\cal A}^0_a(x)$ to define ${\cal A}^\mu(x)$, and $\lambda_a$ are the Gell-Mann matrices in color space. In this work, we consider the magnetic field ${\bf B}=(0, 0, B)$ along the $z$ axis by setting $A_\mu=(0,0,x B,0)$ in Landau gauge, which couples with quarks of electric charge $Q_f=diag(Q_u, Q_d)=diag(2e/3,-e/3)$. $\hat{m}_0=diag(m^u_0, m^d_0)=diag(m_0,m_0)$ is the current quark mass matrix in flavor space, which controls the explicit breaking of chiral symmetry. For the chiral section in the Lagrangian, $G$ is the coupling constant between quarks in the scalar and pseudoscalar channels, which determines the spontaneous breaking of chiral symmetry. The Polyakov potential describing deconfinement at finite temperature reads as
\begin{equation}
\label{polyakov}
{{\cal U}(\Phi,{\bar \Phi})\over T^4} = -{b_2(t)\over 2} \bar\Phi\Phi -{b_3\over 6}\left({\bar\Phi}^3+\Phi^3\right)+{b_4\over 4}\left(\bar\Phi\Phi\right)^2,
\end{equation}
where $\Phi$ is the trace of the Polyakov loop $\Phi=\left({\text {Tr}}_c L \right)/N_c$, with $L({\bf x})={\cal P} \text {exp}[i \int^\beta_0 d \tau {\cal A}_4({\bf x},\tau)]= \text {exp}[i \beta {\cal A}_4 ]$ and $\beta=1/T$, the coefficient $b_2(t)=a_0+a_1 t+a_2 t^2+a_3 t^3$ with $t=T_0/T$ is temperature dependent, and other coefficients $b_3$ and $b_4$ are constants.

The order parameter to describe chiral restoration phase transition is the chiral condensate $\sigma=\langle\bar\psi\psi\rangle$ or the dynamical quark mass $m=m_0-G \sigma $~\cite{njl1,njl2,njl3,njl4,njl5}. $\Phi$ is considered as the order parameter to describe the deconfinement phase transition, which satisfies $\Phi \rightarrow 0$ in confined phase at low temperature and $\Phi \rightarrow 1$ in deconfined phase at high temperature~\cite{pnjl5,pnjl6,pnjl7,pnjl8,pnjl9,pnjl10,pnjl12,t044,t041}. In mean field approximation, the thermodynamic potential at finite temperature, quark chemical potential and magnetic field contains the mean field part and quark part
\begin{eqnarray}
\label{omega1}
&&\Omega(T,\mu,B) ={\cal U}(\Phi,{\bar \Phi})+ \frac{(m-m_0)^2}{2 G}+\Omega_q(T,\mu,B) ,\nonumber\\
&&\Omega_q(T,\mu,B)  = - \sum_{f,n}\alpha_n \int \frac{d p_z}{2\pi} \frac{|Q_f B|}{2\pi} \big[3E_f\\
&& +\ T\ln\left(1+3\Phi e^{-\beta E_f^-}+3{\bar \Phi}e^{-2\beta E_f^-}+e^{-3\beta E_f^-}\right)\nonumber\\
&& +\ T\ln\left(1+3{\bar \Phi} e^{-\beta E_f^+}+3{ \Phi}e^{-2\beta E_f^+}+e^{-3\beta E_f^+}\right)\big],\nonumber
\end{eqnarray}
where $n$ means Landau levels, $f=u,d$ quark flavors, $\alpha_n=2-\delta_{n0}$ spin factor, and $E_f^\pm=E_f \pm \mu_f$ contains quark energy $E_f=\sqrt{p^2_z+2 n |Q_f B|+m^2}$ and quark chemical potential $\mu_u=\frac{\mu_B}{3}+\frac{2\mu_Q}{3}$, $\mu_d=\frac{\mu_B}{3}-\frac{\mu_Q}{3}$ with baryon (electric) chemical potential $\mu_B\ (\mu_Q)$ corresponding to the conserved baryon number (electric charge) $B\ (Q)$.

The ground state at finite temperature, quark chemical potential and magnetic field is determined by minimizing the thermodynamic potential,
\begin{eqnarray}
\label{gapeqs}
\frac{\partial \Omega}{\partial m}=0, \frac{\partial \Omega}{\partial \Phi}=0, \frac{\partial \Omega}{\partial {\bar \Phi}}=0,
\end{eqnarray}
which leads to three coupled gap equations for the order parameters $m$, $\Phi$ and ${\bar \Phi}$. At vanishing quark chemical potential and finite temperature, the chiral symmetry restoration and deconfinement process are smooth crossover. The pseudocritical temperatures $T_{pc}^c$ and $T_{pc}^d$ for chiral restoration and deconfinement phase transitions are determined by the maximum change of the order parameters $\frac{\partial^2 m}{\partial T^2}=0$ and $\frac{\partial^2 \Phi}{\partial T^2}=0$, respectively.

The quadratic fluctuations and correlations of baryon number $B$ and electric charge $Q$ can be obtained by taking the derivatives of the thermodynamic potential $\Omega$ with respect to the chemical potentials $\hat{\mu}_X=\mu_X/T,\ (X=B,\ Q)$ evaluated at zero chemical potential
\begin{eqnarray}
\chi_{i,j}^{B,Q}&=&-\frac{\partial^{i+j}(\Omega/T^4)}{\partial \hat{\mu}_B^i \partial \hat{\mu}_Q^j}{\bigg |}_{{\mu}_X=0},\ i+j=2,
\end{eqnarray}
and the quadratic fluctuation of temperature $T$ is defined as
\begin{eqnarray}
\chi_2^T&=&-\frac{1}{T^2}\frac{\partial^2 \Omega}{\partial T^2}{\bigg |}_{\mu_X=0}.
\end{eqnarray}
In this work, we focus on the correlation $\chi^{BQ}_{11}$ and fluctuations $\chi^B_2,\ \chi^Q_2,\ \chi^T_2$ at finite temperature, magnetic field and vanishing chemical potential.


\section{Numerical Results}
\label{results}
\begin{figure*}[htb]
\begin{center}
\includegraphics[width=7.5cm]{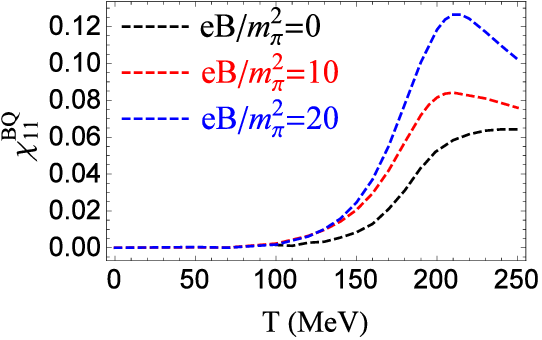}
\includegraphics[width=7.4cm]{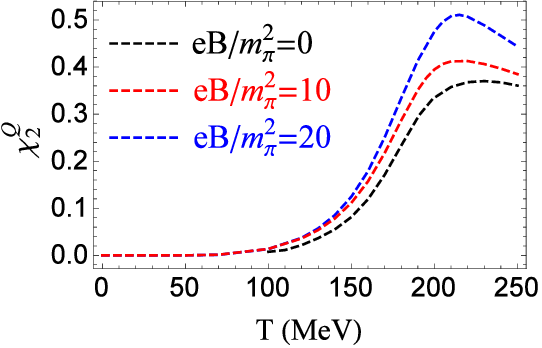}\\
\includegraphics[width=7.5cm]{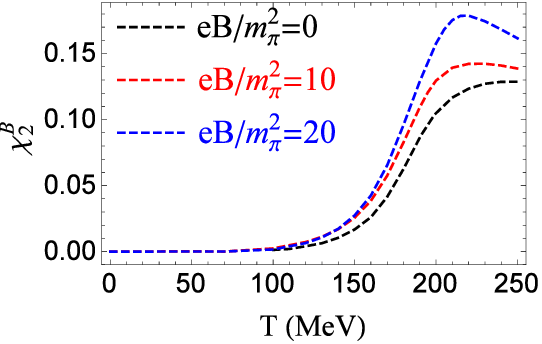}
\includegraphics[width=7.3cm]{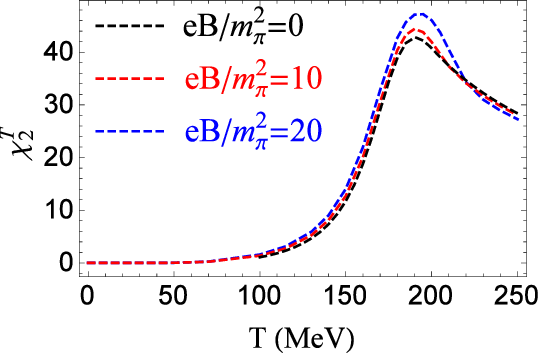}
\end{center}
\caption{Correlation $\chi^{BQ}_{11}$ and quadratic fluctuations $\chi^B_2,\ \chi^Q_2,\ \chi^T_2$ as functions of temperature with fixed magnetic field $eB/m^2_{\pi}=0,\ 10,\ 20$.}
\label{figx2g0t0}
\end{figure*}
\begin{figure}[htb]
\includegraphics[width=7.5cm]{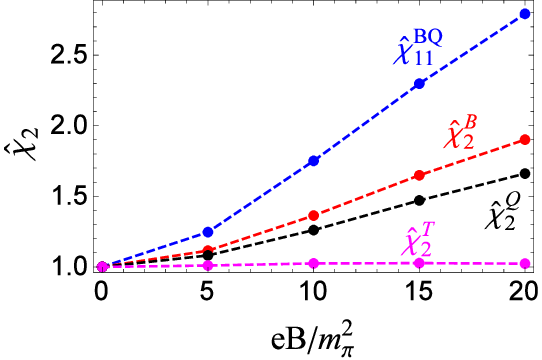}
\caption{The scaled correlation ${\hat {\chi}}_{11}^{BQ}$ and scaled quadratic fluctuations ${\hat {\chi}}_2^{B},\ {\hat {\chi}}_2^{Q},\ {\hat {\chi}}_2^{T}$ at the pseudocritical temperature of chiral restoration phase transition as functions of magnetic field.}
\label{figx2tpcg0t0}
\end{figure}
\begin{figure}[htb]
\centering
\includegraphics[width=7cm]{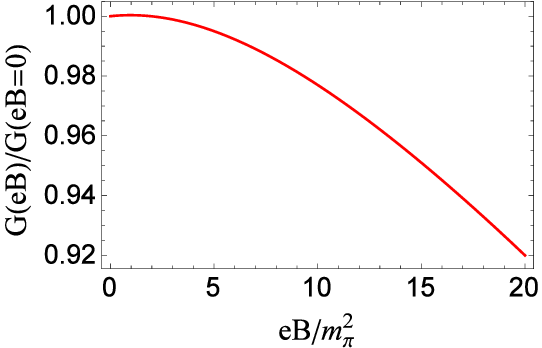}
\includegraphics[width=7cm]{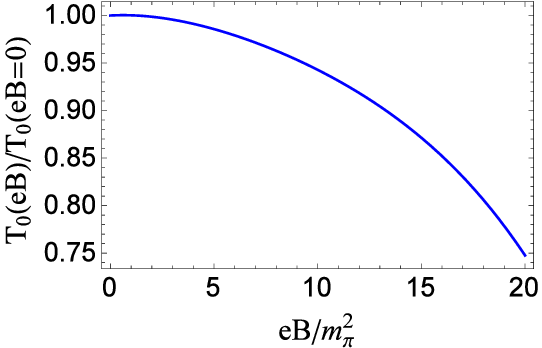}
\caption{Magnetic field dependent parameters $G(eB)$ (upper panel) and $T_0(eB)$ (lower panel) fitted from LQCD reported decreasing pseudocritical temperature of chiral restoration phase transition $T_{pc}^c(eB)/T_{pc}^c(eB=0)$~\cite{lattice1}.} \label{gt0eb}
\end{figure}
\begin{figure*}[htb]
\includegraphics[width=5.8cm]{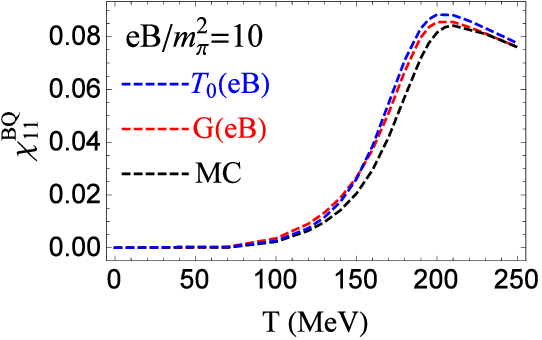}\includegraphics[width=5.8cm]{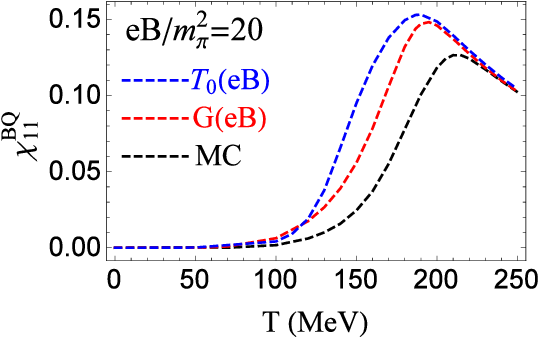}\includegraphics[width=5.6cm]{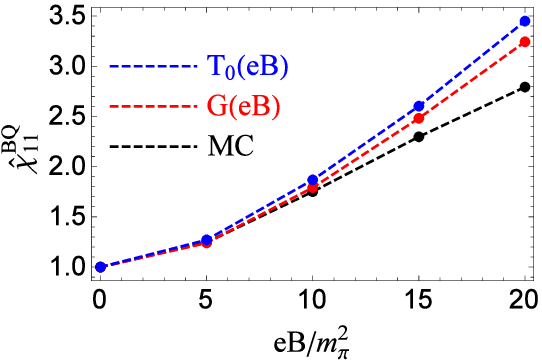}\\
\includegraphics[width=5.8cm]{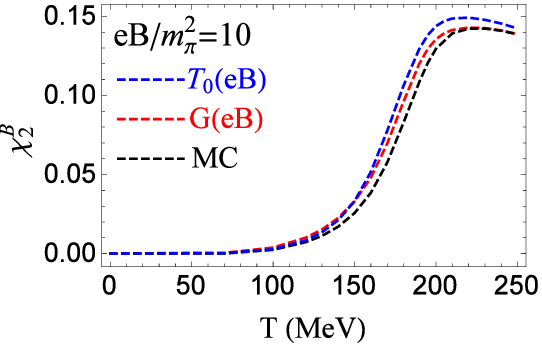}\includegraphics[width=5.8cm]{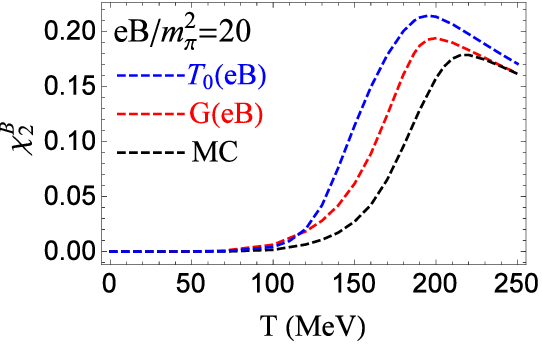}\includegraphics[width=5.6cm]{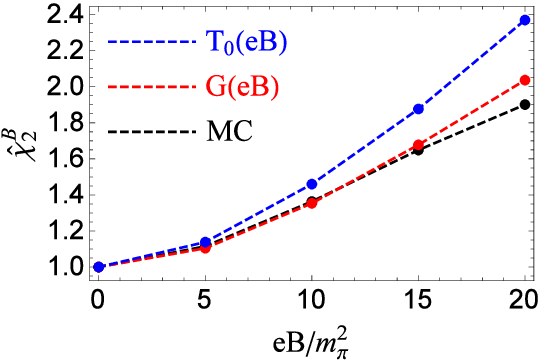}\\
\includegraphics[width=5.8cm]{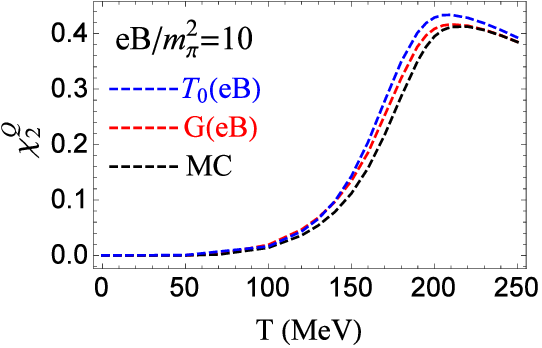}\includegraphics[width=5.8cm]{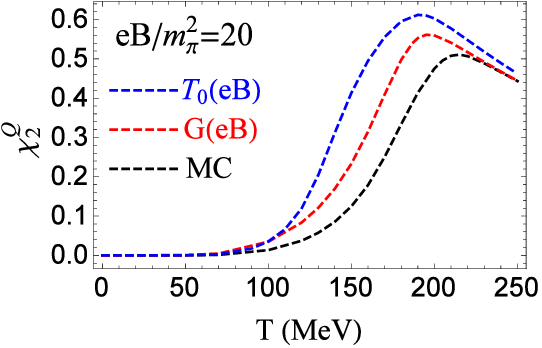}\includegraphics[width=5.7cm]{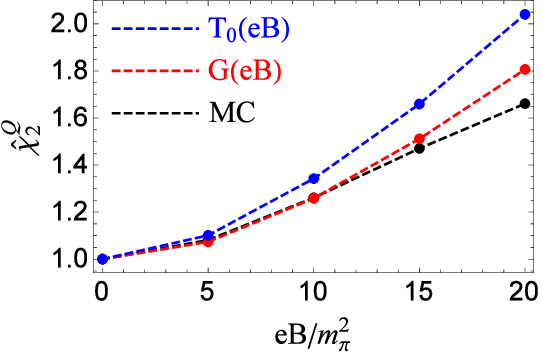}\\
\includegraphics[width=5.7cm]{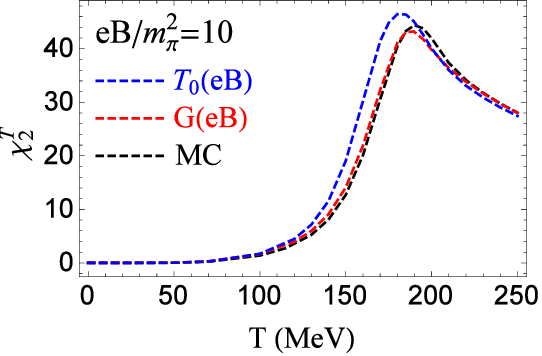}\includegraphics[width=5.7cm]{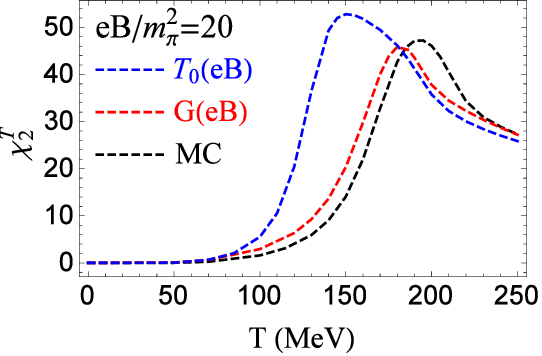}\includegraphics[width=5.8cm]{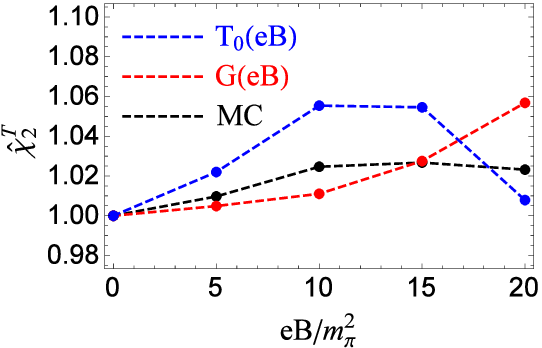}
\caption{(First row) The correlation $\chi^{BQ}_{11}$ as a function of temperature with fixed magnetic field $eB=10m^2_\pi$ (left panel) and $eB=20m^2_\pi$ (middle panel), and the scaled correlation ${\hat{\chi}}^{BQ}_{11}$ (right panel) at $T=T_{pc}^c$ as a function of magnetic field. (Second row) The fluctuation $\chi^{B}_{2}$ as a function of temperature with fixed magnetic field $eB=10m^2_\pi$ (left panel) and $eB=20m^2_\pi$ (middle panel), and the scaled fluctuation ${\hat{\chi}}^{B}_{2}$ (right panel) at $T=T_{pc}^c$ as a function of magnetic field. (Third row) The fluctuation $\chi^{Q}_{2}$ as a function of temperature with fixed magnetic field $eB=10m^2_\pi$ (left panel) and $eB=20m^2_\pi$ (middle panel), and the scaled fluctuation ${\hat{\chi}}^{Q}_{2}$ (right panel) at $T=T_{pc}^c$ as a function of magnetic field. (Forth row) The fluctuation $\chi^{T}_{2}$ as a function of temperature with fixed magnetic field $eB=10m^2_\pi$ (left panel) and $eB=20m^2_\pi$ (middle panel), and the scaled fluctuation ${\hat{\chi}}^{T}_{2}$ (right panel) at $T=T_{pc}^c$ as a function of magnetic field. Here, the blue and red lines correspond to the case with IMC effect by $T_0(eB)$ and $G(eB)$ schemes, respectively, and the black lines correspond to the case without IMC effect.}
\label{figx2gebt0eb}
\end{figure*}

Because of the contact interaction among quarks, NJL models are nonrenormalizable, and it is necessary to introduce a regularization scheme to remove the ultraviolet divergence in momentum integrations. In this work, we take a Pauli-Villars regularization~\cite{mao}, which is gauge invariant and can guarantee the law of causality at finite magnetic field. The three parameters in the two-flavor NJL model, namely the current quark mass $m_0=5$ MeV, the coupling constant $G=7.79$ GeV$^{-2}$, and the mass parameter $\Lambda=1127$ MeV are fixed by fitting the chiral condensate $\langle \bar \psi \psi \rangle =(-250\ \text{MeV})^3$, pion mass $m_{\pi}=134$ MeV, and pion decay constant $f_\pi=93$ MeV in vacuum $(T=\mu=0,\ eB=0)$. For the Polyakov potential, its temperature dependence is fitted from the LQCD simulations, and the parameters are chosen as~\cite{pnjl6,t041,t044} $a_0=6.75$, $a_1=-1.95$, $a_2=2.625$, $a_3=-7.44$, $b_3=0.75$, $b_4=7.5$, and $T_0=210$ MeV.

\subsection{without IMC effect}

The typical results of PNJL model with the original parameters determined from the vacuum properties present the magnetic catalysis effect. The chiral condensates and Polyakov loop increase with the magnetic field in the whole temperature region, and the pseudocritical temepratures for chiral restoration and deconfinement phase transitions also increase with the magnetic field~\cite{db2,dingref35}.

Figure \ref{figx2g0t0} plots correlation $\chi^{BQ}_{11}$ and quadratic fluctuations $\chi^B_2,\ \chi^Q_2,\ \chi^T_2$ as functions of temperature with fixed magnetic field $eB/m^2_{\pi}=0,\ 10,\ 20$. $\chi^{BQ}_{11},\ \chi^B_2$, and $\chi^Q_2$ show similar properties. At vanishing magnetic field, they increase with temperature. Turning on the magnetic field, they have the higher values and the peak structure becomes more pronounced. These properties are consistent with LQCD simulations~\cite{ding2022}. Comparing with the results of $\chi^{BQ}_{11},\ \chi^B_2,\ \chi^Q_2$ in Ref~\cite{dingref35}, where peak structure is observed in $\chi^{BQ}_{11}$ only, the difference is owing to the different regularization schemes, where soft cutoff regularization is operated in the vacuum term $(T=\mu=0)$ of thermodynamical potential in Ref~\cite{dingref35}, and Pauli-Villars regularization is applied in both the vacuum and medium term of thermodynamical potential in our current work. $\chi^T_2$ shows peak structure at vanishing and nonvanishing magnetic fields. The value of $\chi^T_2$ increases with magnetic fields in the region of low temperature and around the peak, but decreases with magnetic fields at high enough temperature.

What is the property of correlation $\chi^{BQ}_{11}$ and fluctuations $\chi^B_2,\ \chi^Q_2,\ \chi^T_2$ along the phase transition line under external magnetic field? As shown in Fig.\ref{figx2tpcg0t0}, the scaled correlation ${\hat {\chi}}_{11}^{BQ}=\frac{\chi_{11}^{BQ}(eB,T_{pc}^c(eB))}{\chi_{11}^{BQ}(eB=0,T_{pc}^c(eB=0))}$ and scaled quadratic fluctuations ${\hat {\chi}}_2^{B(Q,T)}=\frac{\chi_2^{B(Q,T)}(eB,T_{pc}^c(eB))}{\chi_2^{B(Q,T)}(eB=0,T_{pc}^c(eB=0))}$ at the pseudocritical temperature $T_{pc}^c$ of chiral restoration phase transition are plotted as functions of magnetic field. ${\hat {\chi}}^{BQ}_{11}$ increases fastest and it reaches the value of $1.75$ at $eB=10m^2_{\pi}$. LQCD~\cite{dingprl2024} reports ${\hat {\chi}}^{BQ}_{11} \simeq 2.4$ at $eB=8m^2_{\pi}$, which is higher than our PNJL result. ${\hat {\chi}}^{B}_{2}$ monotonically goes up with magnetic fields and approaches the value of $1.36$ at $eB=10m^2_{\pi}$, which undershoots the LQCD result~\cite{dingprl2024}. ${\hat {\chi}}^{Q}_{2}$ monotonically increases with magnetic fields and approaches the value of $1.26$ at $eB=10m^2_{\pi}$, which overshoots the LQCD result~\cite{dingprl2024}. Comparing with Ref\cite{dingref35}, the results of ${\hat {\chi}}^{BQ}_{11},\ {\hat {\chi}}^B_2,\ {\hat {\chi}}^Q_2$ are consistent with Fig.\ref{figx2tpcg0t0}, and not dependent on the regularization schemes. Different from ${\hat {\chi}}^{BQ}_{11},\ {\hat {\chi}}^B_2$, and ${\hat {\chi}}^Q_2$, ${\hat {\chi}}^{T}_{2}$ is not sensitive to the magnetic field. It slightly increases and then slightly decreases with the magnetic field.  \\

\subsection{with IMC effect}

From LQCD simulations~\cite{lattice1,lattice2,lattice4,lattice5,lattice6,lattice7,lattice9}, the IMC phenomenon of chiral symmetry restoration of u and d quarks can be characterized either by the chiral condensates (order parameters) or by the pseudocritical temperature of chiral restoration phase transition. To include the IMC effect in the effect model, one approach is to fit the LQCD results of chiral condensates~\cite{geb1,meson,geb3,geb4}, and another approach is to fit the LQCD results of pseudocritical temperature~\cite{bf8,bf9,geb1,su3meson4,mao11}. The two methods give consistent results with each other. In our current calculations, by fitting the LQCD reported decreasing pseudocritical temperature of chiral symmetry restoration $T_{pc}^c(eB)/T_{pc}^c(eB=0)$ under external magnetic field~\cite{lattice1}, we introduce the IMC effect in the two-flavor PNJL model through a magnetic field dependent parameter $G(eB)$ and $T_0(eB)$, respectively.

On one side, the coupling between quarks plays a significant role in determining the spontaneous breaking and restoration of chiral symmetry. A magnetic field dependent coupling $G(eB)$~\cite{bf8,bf9,geb1,su3meson4,mao11} is introduced into the PNJL model to mimic the reduction of pseudocritical temperature of chiral restoration phase transition under external magnetic field in LQCD calculations~\cite{lattice1}. As plotted in Fig.\ref{gt0eb} upper panel, the magnetic field dependent coupling $G(eB)/G(eB=0)$ is a monotonic decreasing function of magnetic field. On the other side, when considering the magnetic field dependence of the interaction between quarks and Polyakvo loop, a magnetic field dependent parameter $T_0(eB)$ in the Polyakov potential~\cite{t0effectmao,t0effect,pnjl3} is used in the PNJL model to simulate the reduction of pseudocritical temperature of chiral restoration phase transition in LQCD calculations~\cite{lattice1}. As shown in Fig.\ref{gt0eb} lower panel, $T_0(eB)$ is also a monotonic decreasing function of magnetic field. We have checked that, with our fitted parameter $G(eB)$ or $T_0(eB)$, the increase (decrease) of chiral condensates with magnetic fields at the low (high) temperature, the increase of Polyakov loop with magnetic fields in the whole temperature region and the reduction of pseudocritical temperature of deconfinement phase transition under magnetic fields can be realized.

In the following calculations, we use the two schemes $G(eB)$ and $T_0(eB)$ to consider the IMC effect, respectively, and make comparison between the results with and without IMC effect.

Figure \ref{figx2gebt0eb} first three rows depict the results of the correlation $\chi^{BQ}_{11}$, and quadratic fluctuations $\chi^B_2,\ \chi^Q_2$ with IMC effect (blue and red lines) and without IMC effect (black line). At fixed magnetic fields $eB/m^2_\pi=10,\ 20$, when considering the IMC effect with $G(eB)$ and $T_0(eB)$ schemes, $\chi^{BQ}_{11},\ \chi^B_2$, and $\chi^Q_2$ increase with temperature and have the peak around the pseudocritical temperatures of chiral restoration and deconfinement phase transitions, which are similar as the case without IMC effect, see the left and middle panels. And the values of $\chi^{BQ}_{11},\ \chi^B_2,\ \chi^Q_2$ with IMC effect are larger than those without IMC effect in the whole temperature region, although the locations of their peaks are shifted to lower temperatures by the IMC effect. In the right panels, we plot the scaled correlation ${\hat {\chi}}_{11}^{BQ}=\frac{\chi_{11}^{BQ}(eB,T_{pc}^c(eB))}{\chi_{11}^{BQ}(eB=0,T_{pc}^c(eB=0))}$ and scaled quadratic fluctuations ${\hat {\chi}}_2^{B(Q)}=\frac{\chi_2^{B(Q)}(eB,T_{pc}^c(eB))}{\chi_2^{B(Q)}(eB=0,T_{pc}^c(eB=0))}$ at the pseudocritical temperature of chiral restoration phase transition as functions of magnetic field with IMC effect (blue and red lines) and without IMC effect (black line). Including the IMC effect, ${\hat {\chi}}^{BQ}_{11},\ {\hat {\chi}}^B_2$, and ${\hat {\chi}}^Q_2$ increase faster. The results of ${\hat {\chi}}^{BQ}_{11}$ and ${\hat {\chi}}^B_2$ show better quantitative consistency with the LQCD results~\cite{dingprl2024}, but the results of ${\hat {\chi}}^Q_2$ shows larger deviation from the LQCD results~\cite{dingprl2024}. Different methods to consider IMC effect do not lead to qualitative difference in the results, and only cause some quantitative modifications.

%


Figure \ref{figx2gebt0eb} bottom row shows the results of fluctuation $\chi^T_2$ with IMC effect (blue and red lines) and without IMC effect (black line), where the left and middle panels plot $\chi^T_2$ at fixed magnetic field $eB/m^2_\pi=10,\ 20$ as a function of temperature, and the right panel plots the scaled fluctuation ${\hat {\chi}}^T_2=\frac{\chi^T_2(eB,T_{pc}^c(eB))}{\chi^T_2(eB=0,T_{pc}^c(eB=0))}$ at the pseudocritical temperature of chiral restoration phase transition as a function of magnetic field. With fixed magnetic fields, the including of IMC effect shifts the location of the peak of $\chi^T_2$ to a lower temperature. However, the value of $\chi^T_2$ around the peak is enhanced (reduced) when we consider the IMC effect by the method of $T_0(eB)$ ($G(eB)$). The scaled fluctuation ${\hat {\chi}}^T_2$ slightly increases and then decreases as external magnetic field grows in the case without IMC effect (black line). When considering IMC effect by scheme $G(eB)$, ${\hat {\chi}}^T_2$ monotonically increases with magnetic field (red line), but by scheme $T_0(eB)$, ${\hat {\chi}}^T_2$ is a nonmonotonic function of magnetic field (blue line). The introduction of IMC effect causes some qualitative changes in ${\hat {\chi}}^T_2$, which is different from the situations of ${\hat {\chi}}^{BQ}_{11},\ {\hat {\chi}}^B_2$, and ${\hat {\chi}}^Q_2$.

Before the end of this section, we need to make some comments. The LQCD calculation in Ref~\cite{dingprl2024} reports a constant pseudocritical temperature for chiral restoration phase transition under external magnetic field. Such a result can be realized by introducing a magnetic field dependent parameter $G(eB)$ or $T_0(eB)$, which will be different from the results in Fig.\ref{gt0eb}. According to our numerical calculations, this only causes tiny difference for the scaled correlation ${\hat \chi}^{BQ}_{11}$ and scaled fluctuations ${\hat \chi}^B_2,\ {\hat \chi}^Q_2,\ {\hat \chi}^T_2$ at the phase transition.\\

\section{summary}
\label{summary}
The correlation $\chi^{BQ}_{11}$ and fluctuations $\chi^B_2,\ \chi^Q_2,\ \chi^T_2$ at finite temperature and magnetic field are investigated in frame of a two-flavor PNJL model. The IMC effect is introduced into the PNJL model by two methods, the magnetic field dependent coupling between quarks $G(eB)$, and magnetic field dependent interaction between quarks and Polyakov loop $T_0(eB)$. We make comparison of the results in the cases with and without IMC effect.

$\chi^{BQ}_{11},\ \chi^B_2$, and $\chi^Q_2$ show similar properties at finite temperature and magnetic field. With vanishing magnetic field, the correlation $\chi^{BQ}_{11}$ and fluctuations $\chi^B_2,\ \chi^Q_2$ increase with temperature, and with nonvanishing magnetic field, they show the peak around the pseudocritical temperatures of chiral restoration and deconfinement phase transitions. These properties are consistent with LQCD results. To demonstrate $\chi^{BQ}_{11},\ \chi^B_2$, and $\chi^Q_2$ along the phase transition line under external magnetic field, we calculate the scaled correlation ${\hat \chi}^{BQ}_{11}$ and scaled fluctuations ${\hat \chi}^B_2,\ {\hat \chi}^Q_2$ at the pseudocritical temperature $T_{pc}^c$ of chiral restoration phase transition and they increase with magnetic fields. The results of ${\hat \chi}^{BQ}_{11}$ and ${\hat \chi}^B_2$ are qualitatively consistent with LQCD results, but ${\hat \chi}^Q_2$ deviates from LQCD results. The inclusion of IMC effect in the PNJL model does not lead to qualitative difference in the results, and only causes some quantitative modifications.

The fluctuation $\chi^T_2$ presents different properties from $\chi^{BQ}_{11},\ \chi^B_2$ and $\chi^Q_2$. With vanishing and nonvanishing magnetic field, the fluctuation $\chi^T_2$ shows the peak around the pseudocritical temperatures of chiral restoration and deconfinement phase transitions, which is not altered qualitatively by the IMC effect. But the scaled fluctuation ${\hat \chi}^T_2$ at $T=T_{pc}^c$ is modified by the IMC effect. Without IMC effect, ${\hat \chi}^T_2$ slightly increases and then slightly decreases with magnetic fields. Taking into account of the IMC effect by $G(eB)$, ${\hat \chi}^T_2$ monotonically increases with magnetic fields, and by $T_0(eB)$, it is a nonmonotonic function of magnetic field.

As we know, the parameter $T_0=270$ MeV in the Polyakov potential is the critical temperature of deconfinement phase transition in the pure gauge model~\cite{pnjl6}. The inclusion of dynamical quarks leads to a decrease of $T_0$~\cite{t041,t044}. In two flavor case, $T_0=210$ MeV is widely used. It should be mentioned that different choice of parameter $T_0$ will change the pseudocritical temperatures of chiral restoration and deconfinement phase transitions, but as we have checked, it scarcely influences the results of the scaled correlations and fluctuations at the phase transition.

People may wonder why we calculate the correlation $\chi^{BQ}_{11}$ and fluctuations $\chi^{B,Q,T}_2$ in a two-flavor PNJL model, which lacks the component of strangeness. In the temperature region interested here (MeV $100 \leq T \leq 250$ MeV), the strange quark is much more heavier than up and down quarks~\cite{3njlrehberg,3pnjlmei}, and its contribution to the thermodynamical potential and $\chi^{BQ}_{11},\ \chi^{B,Q,T}_2$ can be neglected when comparing with the up and down quarks. Therefore, two-flavor PNJL model describes well the thermodynamics of chiral restoration and deconfinement phase transitions, and it is reliable to make comparison between our results $(\chi^{BQ}_{11},\ \chi^{B,Q,T}_2 )$ and the LQCD results of three-flavor. It should be mentioned that when investigating the correlation and fluctuations related with the strangeness number, we have to apply the three-flavor PNJL model, which is under progress and will be reported in the near future. \\

\noindent {\bf Acknowledgement:} The work is supported by the NSFC grant 12275204.


\begin{thebibliography}{99}

\bibitem{review1} F.Preis, A.Rebhan and A.Schmitt, Lect. Notes Phys. {\bf 871}, 51(2013).
\bibitem{review2} R.Gatto and M.Ruggieri, Lect. Notes Phys. {\bf 871}, 87(2013).
\bibitem{review4} V.A.Miransky and I.A.Shovkovy, Phys. Rep. {\bf 576}, 1(2015).
\bibitem{review5} J.O.Anderson and W.R.Naylor, Rev. Mod. Phys. {\bf 88}, 025001(2016).
\bibitem{review0} G.Q.Cao, Eur. Phys. J. {\bf A 57}, 264 (2021).

\bibitem{review3} M.D'Elia, Lect. Notes Phys. {\bf 871}, 181(2013).
\bibitem{lattice1} G.S.Bali, F.Bruckmann, G.Endrodi, Z.Fodor, S.D.Katz, S.Krieg, A.Schaefer and K.K.Szabo, J. High Energy Phys. {\bf 02}, 044(2012).
\bibitem{lattice2} G.S.Bali, F.Bruckmann, G.Endrodi, Z.Fodor, S.D.Katz and A.Schaefer, Phys. Rev. {\bf D86}, 071502(2012); J. High Energy Phys. {\bf 08}, 177(2014).


\bibitem{lattice9} F.Bruckmann, G.Endrodi and T.G.Kovacs, J. High Energy Phys. {\bf 04}, 112(2013).
\bibitem{lattice4} V.G.Bornyakov, P.V.Buividovich, N.Cundy, O.A.Kochetkov and A.Schaefer, Phys. Rev. {\bf D90}, 034501(2014).
\bibitem{lattice5} G.Endrodi, J. High Energy Phys. {\bf 07}, 173(2015).
\bibitem{lattice6} G.Endrodi, M.Giordano, S.D.Katz, T.G.Kovacs and F.Pittler, J. High Energy Phys. {\bf 07}, 009(2019).
\bibitem{lattice7} H.T.Ding, S.T.Li, J.H.Liu and X.D.Wang, Phys. Rev. {\bf D 105}, 034514(2022).

\bibitem{lattice8} M.D'Elia, F.Manigrasso, F.Negro and F.Sanfilippo, Phys. Rev. {\bf D98}, 054509(2018).


\bibitem{fukushima} K.Fukushima and Y.Hidaka, Phys. Rev. Lett {\bf 110}, 031601(2013).
\bibitem{mao} S.J.Mao, Phys. Lett. {\bf B758}, 195(2016); Phys. Rev. {\bf D94}, 036007(2016); Phys. Rev. {\bf D97}, 011501(R)(2018); Chin. Phys. {\bf C45}, 021004(2021); Phys. Rev. {\bf D106}, 034018(2022).
\bibitem{kamikado}  K.Kamikado and T.Kanazawa, J. High Energy Phys. {\bf 03}, 009(2014).
\bibitem{bf1} J.Y.Chao, P.C.Chu and M.Huang, Phys. Rev. {\bf D88}, 054009(2013).

\bibitem{bf13} J.Braun, W.A.Mian and S.Rechenberger, Phys. Lett. {\bf B755}, 265(2016).
\bibitem{bf2} N.Mueller and J.M.Pawlowski,  Phys. Rev. {\bf D91}, 116010(2015).
\bibitem{bf3} T.Kojo and N.Su, Phys. Lett. {\bf B720}, 192(2013).
\bibitem{bf5} A.Ayala, M.Loewe, A.J.Mizher and R.Zamora, Phys. Rev. {\bf D90}, 036001(2014).
\bibitem{bf51}A.Ayala, L.A.Hernandez, A.J.Mizher, J.C.Rojas and C.Villavicencio, Phys. Rev. {\bf D89}, 116017(2014).
\bibitem{bf52}A.Ayala, C.A.Dominguez, L.A.Hernandez, M.Loewe and R.Zamora, Phys. Rev. {\bf D92}, 096011(2015).
\bibitem{bf8} R.L.S.Farias, K.P.Gomes, G.Krein, and M.B.Pinto, Phys. Rev. {\bf C90}, 025203(2014).
\bibitem{bf9} M.Ferreira, P.Costa, O.Lourenco, T.Frederico, and C.Provid$\hat e$ncia, Phys. Rev. {\bf D89}, 116011(2014).
\bibitem{bf11} F.Preis, A.Rebhan and A.Schmitt, J. High Energy Phys. {\bf 1103}, 033(2011).
\bibitem{db1} E.S.Fraga and A.J.Mizher, Phys. Rev. {\bf D78}, 025016(2008); Nucl. Phys. {\bf A820}, 103C(2009).
\bibitem{db2} K.Fukushima, M.Ruggieri and R.Gatto, Phys. Rev. {\bf D81}, 114031(2010).
\bibitem{db3} C.V.Johnson and A.Kundu, J. High Energy Phys. {\bf 12}, 053(2008).
\bibitem{db5} V.Skokov, Phys. Rev. {\bf D85}, 034026(2012).
\bibitem{db6} E.S.Fraga, J.Noronha and L.F.Palhares, Phys. Rev. {\bf D87}, 114014(2013).
\bibitem{pnjl1} R.Gatto and M.Ruggieri, Phys. Rev. {\bf D82}, 054027(2010), {\bf D83}, 034016(2011).
\bibitem{pnjl2} M.Ferreira, P.Costa and C.Provid$\hat e$ncia, Phys. Rev. {\bf D89}, 036006(2014).
\bibitem{pnjl3} M.Ferreira, P.Costa, D.P.Menezes, C.Provid$\hat e$ncia and N.N.Scoccola, Phys. Rev. {\bf D89}, 016002(2014).
\bibitem{pnjl4} P.Costa,  M.Ferreira,  H.Hansen, D.P.Menezes and C.Provid$\hat e$ncia, Phys. Rev. {\bf D89}, 056013(2014).
\bibitem{pqm} A.J.Mizher, M.N.Chernodub and E.S.Fraga, Phys. Rev. {\bf D82}, 105016(2010).
\bibitem{t0effect} E.S.Fraga, B.W. Mintz and J.Schaffner-Bielich, Phys. Lett {\bf B731}, 154-158(2014).
\bibitem{ferr1} E.J.Ferrer, V.de la Incera, I.Portillo and M.Quiroz, Phys. Rev. {\bf D89}, 085034(2014).
\bibitem{ferr2} E.J.Ferrer, V.de la Incera, and X.J.Wen, Phys. Rev. {\bf D91}, 054006(2015).
\bibitem{meimao1} J.Mei and S.J.Mao, Phys. Rev. {\bf D102}, 114035(2020).
\bibitem{mhuang} K.Xu, J.Y.Chao and M.Huang, Phys. Rev. {\bf D103}, 076015(2021).
\bibitem{meihuangmao} J.Mei, R.Wen, S.J.Mao, M.Huang and K.Xu, Phys. Rev. {\bf D110}, 034024(2024).
\bibitem{t0effectmao} S.J.Mao, Phys. Rev. {\bf D110}, 054002 (2024).


\bibitem{landau} L.D.Landau and E.M.Lifshitz, Quantum Mechanics (Elservier Butterworth-Heinemann, London, 1938).

\bibitem{xwork24} H.T.Ding, F.Karsch and S.Mukherjee, Int. J. Mod. Phys. {\bf E24}, 1530007(2015).
\bibitem{xwork25} W.J.Fu, Commun. Theor. Phys. {\bf 74}, 097304 (2022).
\bibitem{xwork26} X.Luo and N.Xu, Nucl. Sci. Tech. {\bf 28}, 112(2017).
\bibitem{xwork27} A.Pandav, D.Mallic and B.Mohanty, Prog. Part. Nucl. Phys. {\bf 125}, 103960(2022).
\bibitem{xwork28} A.Rustanmov, EPJ Web Conf. {\bf 276}, 01007(2023).
\bibitem{xwork29} T. Nonaka, Acta Phys. Pol. B Proc. Suppl. {\bf 16}, 1(2023).
\bibitem{xwork30} H.S.Ko(STAR  Collaboration), Acta Phys. Pol. B Proc. Suppl. {\bf 16}, 1(2023).

\bibitem{dingref31}K.Fukushima and Y.Hidaka, Phys. Rev. Lett. {\bf 117}, 102301(2016).
\bibitem{dingref33}A.Bhattacharyya, S.K.Ghosh, R.Ray, and S.Samanta, Eur. Phys. Lett. {\bf 115}, 62003(2016).
\bibitem{dingref34}G.Kadam, S.Pal, and A.Bhattacharyya, J. Phys. G {\bf 47}, 125106(2020).
\bibitem{dingref35}W.J.Fu, Phys. Rev. {\bf D88}, 014009(2013).
\bibitem{dingref36}N.Chahal, S.Dutt, and A.Kumar, Phys. Rev. {\bf C107}, 045203(2023).

\bibitem{ding2022} H.T.Ding, S.T.Li, J.H.Liu and X.D.Wang, arXiv:2208.07285.
\bibitem{dingprl2024}H.T.Ding, J.B.Gu, A.Kumar, S.T.Li and J.H.Liu, Phys. Rev. Lett. {\bf 132}, 201903(2024).

\bibitem{pnjl5} P.N.Meisinger and M.C.Ogilvie, Phys. Lett. {\bf B379}, 163(1996).
\bibitem{pnjl6} P.N.Meisinger, T.R.Miller and M.C.Ogilvie, Phys. Rev. {\bf D65}, 034009(2002).
\bibitem{pnjl7} K.Fukushima, Phys. Lett. {\bf B591}, 277(2004).
\bibitem{pnjl8} A.Mocsy, F.Sannino, and K.Tuominen, Phys. Rev. Lett. {\bf 92}, 182302(2004).
\bibitem{pnjl9} E.Megias, E.Ruiz Arriola, and L.L.Salcedo, Phys. Rev. {\bf D74}, 065005(2006).
\bibitem{pnjl10} C.Ratti, M.A.Thaler, and W.Weise, Phys. Rev. {\bf D73}, 014019(2006); arXiv: nucl-th/0604025.

\bibitem{pnjl12} S.K.Ghosh, T.K.Mukherjee, M.G.Mustafa and R.Ray, Phys. Rev. {\bf D73}, 114007(2006).

\bibitem{t041} B.Schaefer, J.M.Pawlowski and J.Wambach, Phys. Rev. {\bf D75}, 074023(2007).
\bibitem{t044} S.Roessner, T.Hell, C.Ratti and W.Seise, Nucl. Phys. {\bf A814}, 118-143(2008).

\bibitem{njl1} Y.Nambu and G.Jona-Lasinio, Phys. Rev. {\bf 122}, 345(1961) and {\bf 124}, 246(1961).
\bibitem{njl2} S.P.Klevansky, Rev. Mod. Phys. {\bf 64}, 649(1992).
\bibitem{njl3} M.K.Volkov, Phys. Part. Nucl. {\bf 24}, 35(1993).
\bibitem{njl4} T.Hatsuda and T.Kunihiro, Phys. Rep. {\bf 247}, 221(1994).
\bibitem{njl5} M.Buballa, Phys. Rep. {\bf 407}, 205(2005).

\bibitem{geb1} H.Liu, L.Yu, M.Chernodub and M.Huang, Phys. Rev. {\bf D94}, 113006(2016).
\bibitem{geb3} A.Ayala, C.A.Dominguez, L.A.Hern$\acute a$ndez, M.Loewe, A.Raya, J.C.Rojas and C.Villavicenico, Phys. Rev. {\bf D94}, 054019 (2016).
\bibitem{geb4} R.L.S.Farias, V.S.Timoteo, S.S.Avancini, M.B.Pinto and G.Klein, Eur. Phys. J. {\bf A53}, 101(2017).
\bibitem{meson} S.Avancini, R.Farias, M.Pinto, W.Travres and V.Tim$\acute{o}$teo, Phys. Lett. {\bf B767}, 247(2017).

\bibitem{su3meson4} S.S.Avancini, M.Coppola, N.N.Scoccola, and J.C.Sodr\'{e}, Phys. Rev. {\bf D104}, 094040(2021).
\bibitem{mao11} S.J.Mao and Y.M.Tian, Phys. Rev. {\bf D106}, 094017(2022).

\bibitem{3njlrehberg} P.Rehberg, S.P.Klevansky, and J.Huefner, Phys. Rev. {\bf C53}, 410(1996).
\bibitem{3pnjlmei}J.Mei, T.Xia, and S.J.Mao, Phys. Rev. {\bf D107}, 074018(2023).

\end{thebibliography}
\end{document}